\documentclass[preprint,12pt]{elsarticle}

\usepackage{amssymb}
\usepackage{graphicx}
\usepackage{amsfonts}
\usepackage{geometry}
\usepackage{epsfig}
\usepackage{CJK}
\usepackage{amsmath}
\usepackage{graphicx}
\usepackage{epstopdf}
\usepackage{mathrsfs}
\usepackage{bbding}
\usepackage{cases}
\usepackage{subfigure}
\usepackage[colorlinks,
           linkcolor=blue,
           anchorcolor=blue,
           citecolor=blue
           ]{hyperref}
\usepackage{float}

\begin{document}

\begin{frontmatter}



\title{Rogue wave and a pair of resonance stripe solitons to a reduced generalized (3+1)-dimensional KP equation}


\author{Xiaoen Zhang,
        Yong Chen$^*$,
        Xiaoyan Tang}

\address{Shanghai Key Laboratory of Trustworthy Computing, East China Normal University, Shanghai, 200062, China}

\begin{abstract}
Based on the bilinear operator and symbol calculation, some lump solutions are presented, rationally localized in all directions in the space, to a reduced (3+1)-dimensional KP equation. The lump solutions all contain six parameters, four of which must cater to the non-zero conditions so as to insure the analyticity and rational localization, while the others are free. Then the interaction between lump soliton and one stripe soliton is described and the result shows that the lump soliton will be drowned or swallowed by the stripe soliton. Furthermore, we extend this method to a new combination of positive quadratic function and hyperbolic functions. Especially, it is interesting that a rogue wave is found to be aroused by the interaction between lump soliton and a pair of resonance stripe solitons. By choosing the values of the parameters, the dynamic properties of lump solution, interaction between lump soliton and one stripe soliton, rogue wave, generated by the interaction between lump soliton and a pair of resonance solitons, are shown graphically.
\end{abstract}


\end{frontmatter}


\section{Introduction}
\label{introduction}
In soliton theory, the study to the integrability of nonlinear equation is always a hot topic. As to the integrable equations, there are many methods to study their solutions, such as the classic inverse scattering method\cite{1}, B$\ddot{a}$cklund transformation\cite{2,3}, Darboux transformation\cite{4}, Hirota bilinear methods\cite{5,6,7}, and variable separation approaches\cite{8,9,10}. Among these methods, the Hirota bilinear method is widely popular due to its  simplicity and directness. Recently, rogue wave solution\cite{11}(being as a special solution of the rational solution) attracts a lot of attention, which was first used to describe the momentous disastrous ocean waves. Its lethality is very strong and can lead to devastating impact to the navigation. There are many ways to get the rogue wave solution, such as generalized Darboux transformation\cite{12}, bilinear method\cite{13,14}, and so on. In contract to the rogue wave solution, lump solution is a special kind of rational solution, rationally localized in all directions in the space. In 2002, Lou et.al studied the lump solution with the variable separation method\cite{15}. Recently, Ma proposed the positive quadratic function to get the lump solution. Special examples of lump solutions have been found, such as the KPI equation\cite{16,17,18}, BKP equation\cite{19},the p-gKP and p-gBKP equations\cite{20}, Boussinesq equation\cite{21} and so on.

More importantly, it will happen collision among different solitons. There are two kinds of collision, either elastic or inelastic. It is reported that lump solutions will keep their shapes, amplitudes, velocities after the collision with soliton solutions, which means the collision is completely elastic\cite{22}. While many other collisions are completely inelastic, for instance,  Becker et.al studied the inelastic collision of solitary waves in anisotropic
Bose-Einstein condensates\cite{23}, Tan discussed the rational breather wave swallowed by kink wave\cite{24}, Tang showed the lump solution drowned by a stripe solution\cite{25}. On the basis of different conditions, the collision will change essentially.

The main purpose of this paper is to study the lump solution, the interaction of lump soliton and one stripe soliton to a generalized (3+1)-dimensional KP equation\cite{26}
\begin{equation}
(u_t+h_1uu_x+h_2u_{xxx}+h_3u_x)_x+h_4u_{yy}+h_5u_{zz}=0, \label{eq.1}
\end{equation}
when $h_1=-1, h_2=-\frac{1}{3}, h_3=1, h_4=1, h_5=-\frac{2}{3}$. Moreover, we extend this method for a combination of positive quadratic function and hyperbolic cosine, then it appears a strange phenomenon, we recall it rogue wave(also be called ghost soliton)from its mathematics expression, generated by the interaction of lump soliton and a pair of resonance solitons in a evolutionary process.

The structure of this paper is as follows: Based on the bilinear operator, Sec.2 obtains the lump solution with the positive quadratic function. Sec.3 list the lump soliton, one stripe soliton and their interaction by using method of the collection of positive quadratic function and exponential function. In the end, we extend the positive quadratic function to a combination of hyperbolic cosine function, during the evolutionary process, there exists a rogue wave as the motivation of the interaction between lump soliton and a pair of resonance solitons.
\section{Lump solution of a reduced generalized (3+1)-dimensional KP equation}
When $h_1=-1, h_2=-\frac{1}{3}, h_3=1, h_4=1, h_5=-\frac{2}{3},$ Eq.(\ref{eq.1}) becomes
\begin{equation}
(u_t-uu_x-\frac{1}{3}u_{xxx}+u_x)_x+u_{yy}-\frac{2}{3}u_{zz}=0.\label{eq.2}
\end{equation}
With the transformation $u=4(\mbox{ln}f)_{xx}$, its bilinear equation can be presented
\begin{equation}
(D_xD_t-\frac{1}{3}D_x^4+D_x^2+D_y^2-\frac{2}{3}D_z^2)f\cdot f=0,\label{eq.3}
\end{equation}
When $z=x$, Eq. (\ref{eq.3}) becomes the following formula
\begin{equation}
\begin{split}
&(D_xD_t-\frac{1}{3}D_x^4+\frac{1}{3}D_x^2+D_y^2)f\cdot f
\\&=2f_{xt}f-2f_xf_t-\frac{2}{3}f_{xxxx}f+\frac{8}{3}f_{xxx}f_{x}-2f_{xx}^2+\frac{2}{3}f_{xx}f-\frac{2}{3}f_x^2+2f_{yy}f-2f_y^2,\label{eq.5}
\end{split}
\end{equation}
which can be changed KP equation with some variable transformations.

Assume
\begin{equation}\label{eq.4}
f=g^2+h^2+a_9, g=a_1x+a_2y+a_3t+a_4, h=a_5x+a_6y+a_7t+a_8,
\end{equation}
where $a_i, 1\leq i\leq 9$ are parameters to be determined. By substituting $f$ into Eq.(\ref{eq.3}), with a direct calculation, these parameters can be expressed:
\begin{numcases}{}\label{eq.6}
a_{3}=\frac{-a_1^3+(3a_6^2-a_5^2-3a_2^2)a_1-6a_5a_6a_2}{3a_5^2+3a_1^2}, a_{7}=\frac{-a_5^3+(3a_2^2-a_1^2-3a_6^2)a_5-6a_1a_6a_2}{3a_5^2+3a_1^2},\\\nonumber a_{9}=\frac{(a_1^2+a_5^2)^3}{(a_1a_6-a_2a_5)^2},
\end{numcases}
which should be satisfied
\begin{equation}\label{eq.7}
a_1a_5\neq 0,~~~~~ \mbox{and} ~~~~~a_1a_6-a_2a_5\neq 0,
\end{equation}
in order to insure the analytical and positive of $f$, meanwhile,
\begin{equation}\label{eq.8}
\begin{vmatrix}
a_1 &a_2\\
a_5 &a_6
\end{vmatrix}\neq 0
\end{equation}
guarantee the rationally localization of $u$ in all directions in the $(x, y)$-plane.

In return, substitute Eq.(\ref{eq.6}) into Eq.(\ref{eq.4}) and generate a class of positive, analytical $f$:
\begin{equation}
\begin{split}
f&=(a_1x+a_2y+\frac{-a_1^3+(3a_6^2-a_5^2-3a_2^2)a_1-6a_5a_6a_2}{3a_5^2+3a_1^2}t+a_4)^2
+\frac{(a_1^2+a_5^2)^3}{(a_1a_6-a_2a_5)^2}
\\&+(a_5x+a_6y+\frac{-a_5^3+(3a_2^2-a_1^2-3a_6^2)a_5-6a_1a_6a_2}{3a_5^2+3a_1^2}t+a_8)^2,
\end{split}
\end{equation}
then the solution of $u$ can be written through the transformation $u=4(\mbox{ln}f)_{xx}$
\begin{equation}
\begin{split}\label{eq.9}
u&=\frac{8(a_1^2+a_5^2)}{(a_1x+a_2y+a_3t+a_4)^2
+(a_5x+a_6y+a_7t+a_8)^2+a_9}
\\&-\frac{16((a_1x+a_2y+a_3t+a_4)a_1+(a_5x+a_6y+a_7t+a_8)a_5)^2}{((a_1x+a_2y+a_3t+a_4)^2
+(a_5x+a_6y+a_7t+a_8)^2+a_9)^2}.
\end{split}
\end{equation}
In this class of lump solution, Eq.(\ref{eq.7}) and Eq.(\ref{eq.8}) should satisfy, its expicity pictures are showed when $y=0$ and $t=0$ respectively.
Moreover, its density plots are given and the shape of lump solution is clearer.
\section{The interaction between lump soliton and one stripe soliton}
We want to study the collision between the lump soliton and one stripe soliton. In sec.2, based on the positive quadratic function, its lump solution are presented. In this section, make $f$ as a combination of positive quadratic function and one exponential function, that is
\begin{equation}
f_1=m_1^2+n_1^2+l_1+a_9, \label{eq.10}
\end{equation}
where
\begin{equation*}
m_1=a_1x+a_2y+a_3t+a_4, n_1=a_5x+a_6y+a_7t+a_8, l_1=ke^{k_1x+k_2y+k_3t},
\end{equation*}
through substituting Eq.(\ref{eq.10}) into Eq.(\ref{eq.5}) and symbol calculation, these parameters are calculated:
\begin{equation}\label{eq.11}
a_1=-\frac{a_6}{k_1}, a_2=0, a_3=-\frac{a_6(-1+3k_1^2)}{3k_1}, a_5=0, a_7=0, a_9=\frac{a_6^2}{k_1^4}, k_2=0, k_3=\frac{1}{3}(k_1^3-k_1),
\end{equation}
which needs to satisfy conditions
\begin{equation}\label{eq.12}
k_1\neq 0,~~~k>0,
\end{equation}
to make the corresponding solutions $f$ is positive, analytical and guarantee the localization of $u$ in all directions in the $(x,y)$-plane.

Based on the transformation $u=4(\mbox{ln}f_1)_{xx}$, the solution of Eq.(\ref{eq.5}) will be got again
\begin{equation}\label{eq.13}
u=\frac{4(2a_1^2+2a_5^2+k_1^2l)}{f_1}-\frac{4(2a_1m+2a_5n+k_1l)^2}{f_1^2},
\end{equation}
where
\begin{equation*}
\begin{split}
&f_1=(-\frac{a_6x}{k_1}-\frac{a_6(-1+3k_1^2)t}{3k_1}+a_4)^2+(a_6y+a_8)^2+ke^{k_1x+(\frac{k_1^3}{3}-\frac{k_1}{3})t}+\frac{a_6^2}{k_1^4},\\&
m_1=-\frac{a_6x}{k_1}-\frac{a_6(-1+3k_1^2)t}{3k_1}+a_4, n_1=a_6y+a_8, l_1=ke^{k_1x+(\frac{k_1^3}{3}-\frac{k_1}{3})t}.
\end{split}
\end{equation*}
By choosing appropriate values of these parameters, the dynamic graphs of collision between the lump soliton and one stripe soliton are showed in Fig. \ref{fig3}, Fig. \ref{fig4}:
Whereafter, its density plots are shown in Fig. \ref{fig4}
Fig. \ref{fig4} (a) shows there is one lump soliton and one stripe soliton, the energy of lump soliton is stronger than stripe soliton, when $t$ is up to 0, lump soliton begins to be swallowed by stripe soliton step by step, its energy begin to transfer into the stripe soliton gradually, until it is swallowed by the stripe soliton completely, these two kinds of solitons roll into one soliton and continue to spread.
\section{Rogue wave and a pair of resonance solitons}
Based on the collision of lump soliton and one stripe soliton, we begin to discuss the collision of lump soliton and two stripe solitons. Take $f$ as the combination of positive quadratic function and two exponential functions in virtue of method to seek N-soliton solutions of bilinear form, that is
\begin{equation}\label{eq.30}
f_2=m_2^2+n_2^2+kg_2+k_4h_2+k_8g_2h_2,
\end{equation}
where
\begin{equation*}
m_2=a_1x+a_2y+a_3t+a_4, n_2=a_5x+a_6y+a_7t+a_8, g_2=e^{k_1x+k_2y+k_3t}, h_2=e^{k_5x+k_6y+k_7t},
\end{equation*}
by substituting Eq.(\ref{eq.30}) into Eq.(\ref{eq.5}), we can get
\begin{numcases}{}
a_1=\frac{a_6}{k_5}, a_2=0, a_3=\frac{a_6(-1+3k_5^2)}{k_5}, a_5=0, a_7=0, a_9=\frac{a_6^2}{k_5^4}, k=\frac{a_6^2k_8}{k_4k_5^4}, \\\nonumber k_1=-k_5, k_2=0, k_3=\frac{k_1^4-3k_2^2-k_1^2}{3k_1}, k_6=0, k_7=\frac{k_5^4-k_5^2-3k_6^2}{3k_5},
\end{numcases}
the results indicates these two exponential functions are a pair of resonance solitons, by some transformations, it can be changed into a hyperbolic cosine function, hence, reinstall $f$ as the following formula:
\begin{equation}\label{eq.14}
f_3=m_3^2+n_3^2+k\mbox{cosh}(k_1x+k_2y+k_3t)+a_9,
\end{equation}
where
\begin{equation*}
m_3=a_1x+a_2y+a_3t+a_4, n_3=a_5x+a_6y+a_7t+a_8,
\end{equation*}
once again, substitute Eq.(\ref{eq.14}) into Eq.(\ref{eq.5}), with a complex symbol calculation, the relations of these parameters are
\begin{numcases}{}\label{eq.15}
a_1=\frac{a_6}{k_1}, a_3=\frac{a_6^2(-1+3k_1^2)-3a_2^2k_1^2}{3k_1a_6}, a_5=0, a_7=-2a_2k_1, a_9=\frac{k^2k_1^8+4a_6^4}{4a_6^2k_1^4}, \\\nonumber k_2=\frac{a_2k_1^2}{a_6}, k_3=\frac{(k_1^3-k_1)a_6^2-3a_2^2k_1^3}{3a_6^2},
\end{numcases}
which needs to satisfy
\begin{equation}
k_1\neq 0, ~~~a_6\neq 0, ~~~k>0,
\end{equation}
to guarantee the corresponding solutions $f_3$ is positive, analytical and insure the localization of $u$ in all directions in the $(x,y)$-plane.

Still, institute Eq.(\ref{eq.15}) into Eq.(\ref{eq.14}), with the transformation $u=4(\mbox{ln}f_3)_{xx}$, we can obtain the solution of $u$
\begin{equation}\label{eq.16}
\begin{split}
u&=\frac{4(2a_1^2+2a_5^2+k\mbox{cosh}(k_1x+k_2y+k_3t)k_1^2)}{f_3}\\&-4\frac{(2a_1m_3+2a_5n_3+kk_1\mbox{sinh}(k_1x+k_2y+k_3t))^2}{f_3^2},
\end{split}
\end{equation}
where
\begin{numcases}{}
\nonumber
f_3=(\frac{a_6}{k_1}x+a_2y+\frac{a_6^2(-1+3k_1^2)-3a_2^2k_1^2}{k_1a_6}t+a_4)^2+(a_6y-2a_2k_1t+a_8)^2+\frac{k^2k_1^8+4a_6^4}{4a_6^2k_1^4}\\+k\mbox{cosh}(k_1x+\frac{a_2k_1^2}{a_6}y+ \frac{(k_1^3-k_1)a_6^2-3a_2^2k_1^3}{3a_6^2}t),
~~n_3=a_6y-2a_2k_1t+a_8, \\\nonumber m_3=\frac{a_6}{k_1}x+a_2y+\frac{a_6^2(-1+3k_1^2)-3a_2^2k_1^2}{k_1a_6}t+a_4.
\end{numcases}
According to the expression of $f_3, n_3, m_3$, asymptotic property of lump solution and a pair of resonance solitons are analyzed.

Take
\begin{equation*}
\begin{split}
&\xi_{1}=\frac{a_6}{k_1}x+a_2y+\frac{a_6^2(-1+3k_1^2)-3a_2^2k_1^2}{k_1a_6}t+a_4, ~~~\xi_2=a_6y-2a_2k_1t+a_8, \\&\xi_3=k_1x+\frac{a_2k_1^2}{a_6}y+ \frac{(k_1^3-k_1)a_6^2-3a_2^2k_1^3}{3a_6^2}t,
\end{split}
\end{equation*}
with a comparison for $\xi_1, \xi_2, \xi_3$, it is proved
\begin{equation*}
\xi_1=\xi_3\frac{a_6}{k_1^2}+\frac{2a_6k_1t}{3}+a_4, \lim_{t=\pm\infty}\frac{\xi_1^2}{\xi_2^2}=\frac{(3a_2^2k_1^2-3a_6^2k_1^2+a_6^2)^2}{k_1^4a_6^2a_2^2}
\end{equation*}
due to $t=\pm\infty$, $\xi_1, \xi_2$ are same order, we only need to compare $\xi_1^2, \mbox{cosh}\xi_3$, if suppose $\xi_3$ is a constant, then $\xi_1$ is a combination of scale change and time displacement for $\xi_3$. When $t=\pm\infty$,
\begin{equation*}
\lim_{t=\pm\infty}\frac{\xi_1^2}{\mbox{cosh}(\xi_3)}=\lim_{t=\pm\infty}\frac{(\xi_3\frac{a_6}{k_1^2}+\frac{2a_6k_1t}{3}+a_4)^2}{\mbox{cosh}(\xi_3)}=0,
\end{equation*}
so we can come to a conclusion, when $t=\pm\infty$, there only are a pair of resonance solitons, when $t$ becomes little, lump soliton property is more obvious, which can be seen in Fig. \ref{fig6}.
Fig. \ref{fig6} (a) indicates there are a pair of resonance solitons, lump soliton is in a invisible place, similar to a ghoston, (b) shows when $t=-2$, lump soliton appears gradually, it attaches to one of the resonance stripe soliton. More importantly, because of energy conservation, the shapes of these two resonance solitons change at the same time and in the same location, one position appears a lump soliton, the other relative position appears a sunk envelope. When $t=0$, there exists a rogue wave, derived from lump soliton, is located in the middle of these two resonance solitons and link them with each other, then its carrier begin to transfer, until it attaches to other stripe soliton successfully and out of our vision.
Fig. \ref{fig7} (a) shows there are only a pair of resonance stripe solitons, lump soliton maybe as a ghoston, which is hidden in one of the stripe solitons, (b) shows lump soliton appears in one of stripe soliton, when $t$ is up to 0, lump soliton's energy reaches up to the maximum, which presents the property of rogue wave. Whereafter, its energy transfer into the other stripe until disappearing. This whole progress can be regarded as a appearing of the rogue wave. Then its sectional drawing and vertical view are showed in Fig. \ref{fig8} Fig. \ref{fig9} respectively
green line represents t=-10, blue line represents t=10, red line represents t=0. Obviously, the amplitude of t=0 is about five times than t=$\pm 10$ as well as its appearing
time is short, which is cater to the properties of rogue wave. So this kind of lump solution can
also be called rogue wave, which is aroused by the interaction between lump soliton and a pair of resonance solitons.

\section{Discussion}
Based on the Hirota formulation and symbol calculation, we research a reduced (3+1)-dimensional KP equation (\ref{eq.2}). First, its lump soliton is got and the analyticity, localization of the resulting is guaranteed by the non-zero determinant and some constraints to the parameters. But not every Eq.(\ref{eq.1}) have lump soliton, when $h_1=1, h_2=1, h_3=1, h_4=1, h_5=1$, the bilinear formulation of Eq.(\ref{eq.1}) is
\begin{equation}
(D_xD_t+D_x^4+D_x^2+D_y^2+D_z^2)f\cdot f=0.
\end{equation}
Once more, set $z=x$ and assume $f$ as (\ref{eq.4}), then the quadratic function of $f$ is
\begin{equation*}
\begin{split}
&f=(a_1x+a_2y+\frac{-a_1^3+(-2a_5^2-a_2^2+a_6^2)a_1-2a_5a_6a_2}{a_1^2+a_5^2}+a_4)^2+(-\frac{3(a_1^2+a_5^2)^3}{(a_1a_6-a_2a_5)^2})\\
&\hspace{0.5cm}+(a_5x+a_6y+\frac{-a_5^3+(-2a_1^2-a_6^2+a_2^2)a_5-2a_1a_6a_2}{a_1^2+a_5^2})^2,
\end{split}
\end{equation*}
we can see $a_9<0$ in the expression of $f$, so there doesn't exist lump soliton, let alone the collision with other solitons.

Second, we describe the collision in lump soliton and one stripe soliton, the dynamic property is presented in Fig. \ref{fig3}, in the beginning, lump soliton keep its types, energy, spread with a steady rate, but when it meets the stripe soliton, it will interact with the stripe soliton, but lump soliton is swallowed in the end.

Furthermore, we characterize the rogue wave and a pair of resonance solitons, their dynamic property is showed in Fig. \ref{fig6}, in the beginning, there have a pair of resonance stripe solitons, but their amplitudes are different, we can see there should exist a mysterious soliton. Then we discover there appears a small lump soliton which attaches to one of stripe soliton, and it sinks a wave packet in the homologous location for the other stripe soliton. As times goes on, the interaction between this lump soliton and resonance solitons becomes more vigorous, in a special time, its amplitude is up to the maximum, which can be seen in Fig. \ref{fig8}. Its amplitude changes greatly and its occurrence time is short, which caters to the character of rogue wave, so this progress can be called the generation of the rogue wave. Whereafter, rogue wave degenerate into a lump soliton and attaches to the resonance soliton again until disappearing. As we all know, rogue wave was first be found to the KP type equation.

In addition, we should try to discuss the interactions between lump soliton and other kinds of solitons. Moreover, we can expand this method to discrete equations and study their lump solution. These problems will be more interesting and be worthy of discussing.

\section*{Acknowledgment}
We would like to express our sincere thanks to S Y Lou, W X Ma
and other members of our discussion group for their valuable comments.\\
The project is supported by the Global Change Research Program of China(No.2015CB953904), National Natural Science Foundation of China (No. 11275072, 11435005, 11675054, 11275123
 and 11675055). The Network Information Physics Calculation of basic research innovation research group of China (No.61321064), and Shanghai Collaborative Innovation Center of Trustworthy Software for Internet of Things (No.ZF1213).





\end{document}